\documentclass[11pt,a4paper]{article}
\pdfoutput=1
\usepackage{euscript,amsmath,amsthm,amsfonts,amssymb}
\usepackage{mathtools}
\usepackage{subfig}
\usepackage{float}
\usepackage[margin=1in]{geometry}
\usepackage{graphicx}
\usepackage{outlines}
\usepackage[dvipsnames]{xcolor}
\usepackage{jheppub}
\newcommand{\ket}[1]{|{#1}\rangle}

\newcommand{\be}{\begin{equation}}
\newcommand{\ee}{\end{equation}}
\newcommand{\bea}{\begin{eqnarray}}
\newcommand{\eea}{\end{eqnarray}}

\DeclareMathOperator{\area}{area}

\newtheorem{theorem}{Theorem}

\newtheorem{conjecture}{Conjecture}

\title{\boldmath Hyperthreads in holographic spacetimes}
\author{Jonathan Harper}
\affiliation{Martin Fisher School of Physics, Brandeis University, Waltham, Massachusetts 02453, USA}
\preprint{BRX-TH-6688}
\emailAdd{jharper@brandeis.edu}
\abstract{We generalize bit threads to hyperthreads in the context of holographic spacetimes. We define a ``$k$-thread" to be a hyperthread which connects $k$ different boundary regions and posit that it may be considered as a unit of $k$-party entanglement. Using this new object, we show that the contribution of hyperthreads to calculations of holographic entanglement entropy are generically finite. This is accomplished by constructing a surface whose area determines their maximum allowed contribution. We also identify surfaces whose area is proportional to the maximum number of $k$-threads, motivating a possible measure of multipartite entanglement. We use this to make connections to the current understanding of multipartite entanglement in holographic spacetimes.

}
\begin{document}
\maketitle
\flushbottom

\newpage
\section{Introduction}

The connection between geometry and quantum information has a long and rich history in the context of holographic systems. Per the Ryu-Takayanagi (RT) formula \cite{Ryu:2006bv}, the entanglement entropy of a part of a holographic CFT state is given by the area of the minimum surface homologous to the region of interest $A$ in a constant time slice of the dual spacetime  
\be
    S(A)=\min_{m\sim A} \area (m).
\ee
Tools from convex optimization theory \cite{boyd2004convex} allow this problem to be recast as a dual maximization problem which asks for the maximum number of ``bit threads" \cite{Freedman_2016,Headrick_2018} which connect $A$ to its compliment while satisfying a local norm bound
\be
    S(A)= \max_{v^{\mu}}\int_A \sqrt{h} n_{\mu}v^{\mu}, \text{ s.t. } \nabla_{\mu}v^{\mu}=0, \; |v^\mu|\leq 1.
\ee
Here $\sqrt{h}$ is the induced metric on the boundary and $n^\mu$ the unit normal. The integral curves of the vector field $v^{\mu}$ are the individual bit threads which can be thought of as a distilled bell pair of two qubits residing at the endpoints of the thread. The constraints of $v^\mu$ dictate that threads cannot begin or end in the bulk and take up a finite amount of space in the geometry. As such the minimal surface functions as a bottleneck, limiting the number of bit threads permitted through. As a result the maximum number of threads is given precisely by the entanglement entropy as demanded by the duality.

 A better understanding of the role of multipartite entanglement in holographic systems is a longstanding question in the field. We can begin to make progress towards this goal via the bit threads formalism. If a bit thread represents two entangled qubits, then it is natural to conjecture that an object which connects $k$-parties should be suitably thought of as a some unit of $k$-party entanglement between the $k$ boundary regions \cite{bao2020bit}. Such ``hyperthreads" will be the primary focus of this article.

Because hyperthreads do not have a good representation as the integral curves of a vector field we are required to adopt an alternate formulation. Consider the space of all such possible hyperthreads as a measurable set $H$. In the context of this set, we can use tools of measure theory to define the optimization program
\be
\max \sum_{k=2}^nk\int_{H_k}d\mu(h) \text{ s.t. } \forall x\in \Sigma,\; \int_{H}d\mu(h)\Delta(x,h)\leq 1.
\ee
Here $H_{k}\subset H$ is the space of all hyperthreads connecting $k$ different boundary regions (``$k$-threads"), with each term weighted by the number of qubits (endpoints) it contributes. The optimization is over the possible measures $\mu$ of $H$ with the constraint that in the manifold the local density of hyperthreads is at most 1. The measure is essentially an assignment of weights to various configurations of hyperthreads. We will show that this program is dual to
\be
\min \nu(x)\text{ s.t. } \forall h\in H_k,\; \int_{\Sigma}d\nu(x)\Delta(x,h) \geq k
\ee
which asks for the smallest possible configuration or barriers such that any $k$-thread will cross barriers with a total weight of at least $k$. The 2-threads turn out to provide the strictest constraint so that the solution is given precisely by the union of the minimal surfaces associated to the single party entanglement entropies. That is, given a partition of the boundary into $n$ regions
\be
\nu^*=\sum_{k=2}^nk\mu^*(H_k)=\sum_i^n\area(m_{A_i})=\sum_i^n S(A_i),
\ee
where $\nu^*$, $\mu^*$ are the optimal measures. We will further show that the maximum contribution of hyperthreads $k\geq3$ to the entanglement entropy is finite and demonstrate that for AdS$_3$ it is given by the area of a particular surface which partitions the interior region $\Sigma\setminus \cup_i r(A_i)$. Here $r(A_i)$ is the homology region of $A_i$ which interpolates between $A_i$ and $m_{A_i}$. Though the precise consequences of this to the structure of entanglement of CFT states is unclear, this surface could possibly quantify a limitation to the distillation of multipartite entanglement. This would potentially provide another example of holography encoding key properties of quantum information in geometric surfaces.

We will also establish programs which ask for a maximal configuration of \emph{only} $k$-threads for a fixed value of $k$ such that $2<k\leq n$
\be
\begin{split}
HP_{k}(\mathcal{A})&=\max\; k\mu(H_k)\text{ s.t. } \forall x\in \Sigma,\; \int_{H}d\mu(h)\Delta(x,h)\leq 1\\
&=\min \nu(x)\text{ s.t. } \forall h\in H_k,\; \int_{\Sigma}d\nu(x)\Delta(x,h) \geq k.
\end{split}
\ee
Though similar, the lack of 2-threads places strong conditions on the hyperthread configuration making the optimal value substantially smaller. The resulting minimal surface in vacuum AdS$_3$ is given by a configuration which partition the bulk into $k$ regions one of which is homologous to each boundary region. Notably this  allows for bulk intersection points. Here we have defined the $k$-hyperthread partition $HP_{k}(\mathcal{A})$ as the optimal value of this program and conjecture that this might be a valuable measure of $k$-party entanglement. However, to back this proposal, one would desire supporting calculations in the boundary CFT, possibly through replica trick methods.

The plan of this paper is as follows: In section \ref{sec:td} we review bit threads and establish their formulation in the framework of measurable sets\footnote{The measure theory approach to bit threads presented in this paper is based on to-be published work by Matt Headrick and Veronika Hubeny \cite{cbt}. We thank them for sharing an early draft and allowing us to present it here.}. We then state a theorem which helps determine which threads can contribute to an optimal configuration using complementary slackness, a key result from optimization theory. Next in section \ref{sec:htc} we generalize to the case of hyperthreads and show that their contribution to the calculation of the entanglement entropy is finite. This is done by constructing a surface whose area gives the maximum contribution of $k$-threads. We also establish a duality between maximal configurations of $k$-threads and surfaces in the bulk spacetime. Finally, in section \ref{sec:dis} we touch on some additional related topics, including some speculation of the connection of our results to the measurement and distillation of multipartite entanglement in holographic states. 

\section{Bit threads}\label{sec:td}
To begin, we will first review bipartite 2-threads in holographic spacetimes.\footnote{We will assume some basic knowledge of bit threads and convex optimization. For an introduction to convex optimization and applications to bit threads we suggest reviewing sections 2 and 3 of \cite{Headrick_2018}.} Given a holographic spacetime $M$ we take a constant time slice $\Sigma$ and partition the boundary $\partial \Sigma$ into two regions $A$ and $B$. The entanglement entropy of $A$ is given by the area of minimal surface homologous to $A$. This is the Ryu-Takayanagi (RT) formula \cite{Ryu:2006bv}
\be
S(A)=\min_{m\sim A}\area(m)=\area(m_A),
\ee
where we have called the minimizing RT surface $m_A$. Alternatively, using strong duality the RT formula can be written as the maximum flow of bit threads
\be\label{pgm:flow}
    S(A)= \max_{v^{\mu}}\int_A \sqrt{h} n_{\mu}v^{\mu}, \text{ s.t. } \nabla_{\mu}v^{\mu}=0, \; |v^\mu|\leq 1.
\ee
Our goal will be to reinterpret this program as an optimization over a measureable space. We define a thread $p$ to be a simple curve on $\Sigma$ which has one end point on $A$ and one on $B$. We then define the set of all such threads to be $P$.\footnote{As defined $P$ depends on the choice of bipartition of the boundary. Let $P_{all}$ be the set of all possible threads even those which connect $A$ to $A$ and $B$ to $B$. This set does \emph{not} depend on the bipartition. We can always optimize over $P_{all}$ with the additional constraint $\mu(P_{all}\setminus P)=0$. To keep things simple we will work with $P$ with the understanding that this procedure can always be done if desired.} The measure $\mu$ assigns zero or a positive value to subsets of $P$, that is collections of threads. Comparing with the flow program \eqref{pgm:flow} the divergencelessness constraint is automatically implemented since $P$ does not contain any threads with endpoints in the bulk.

In order to impose the norm bound constraint we need to define a local density on threads on $\Sigma$. First we construct a delta function on $P$
\be
\Delta (x,p) \coloneqq \int_p ds\delta(x,p(s))
\ee
where $p$ is a thread and $s$ an affine parameter along it. This counts the number of times, or multiplicity, that the thread intersects with a point $x\in \Sigma$.

From this we have a thread density as a function of $x$ for a given measure $\mu$\footnote{For intuition it is often convenient to consider a measure where the configuration of a single thread is assigned a measure of 1. This allows one to "count" the threads and easily determine how big a configuration is. This is especially useful when considering examples with small numbers of threads such as a graph. However, this measure is incompatible with the density bound as defined since generally a maximal thread configuration will contain an infinite number of threads. For the feasible measures we will encounter (those satisfying the constraints) typically any single thread or finite collection of threads has measure zero. As such, in reality, there is no  notion of building a thread configuration as the union of individual threads.}
\be
\rho(x) \coloneqq\int_{P}d\mu(p)\Delta(x,p).
\ee

We can then use this to define a valid thread configuration as a measure $\mu$ such that
\be
\forall x\in \Sigma,\quad \int_{P}d\mu(p)\Delta(x,p)\leq 1
\ee
which is equivalent to the norm bound constraint in \eqref{pgm:flow}. Putting this together the entanglement entropy can be calculated as a constrained maximization program over the space of measures on $P$:

\be\label{eq:bimax}
S(A)=\max \mu(P)\text{ s.t. } \forall x\in \Sigma,\; \rho(x) \leq 1.
\ee
We will now apply strong duality to the convex optimization program \eqref{eq:bimax}. This is done in two steps: first, we construct a Lagrangian from the objective using Lagrange multipliers to impose the constraints. It is important to note that inequality constraints require the associated Lagrange multiplier to be positive.
\be
\begin{split}
L(d\mu,d\nu)&=\int_{P}d\mu(p)-\int_{\Sigma}d\nu(x)\left(\int_{P}\left(d\mu(p)\Delta(x,p)\right)- 1\right)\\
&=\int_{P}d\mu(p) \left(1-\int_{\Sigma}d\nu(x)\Delta(x,p)\right)+\int_{\Sigma}d\nu(x).
\end{split}
\ee
Next, we optimize with respect to the original variables $d\mu$. This gives us the dual minimization program in terms of the original Lagrange multipliers $d\nu$
\be\label{eq:bimin}
\min \nu(\Sigma)\text{ s.t. } \forall p\in P,\; \int_{\Sigma}d\nu(x)\Delta(x,p) \geq 1.
\ee
We can think of the measure $\nu$ as a collection of level sets. The constraint requires that each thread $p\in P$ must intersect at least $1$ unit of level sets. Since the objective is minimized an optimal configuration will consist of all level sets being placed on the smallest possible surface which satisfies the constraints, as placing a level set elsewhere would only increase the objective. Thus, the level sets accumulate on the minimal surface homologous to $A$ forming a barrier that every thread crosses. The homology condition is naturally imposed since \emph{all} 2-threads in $P$ will cross $m_A$.
\begin{figure}[H]
\centering
\includegraphics[width=.5\textwidth,page=2]{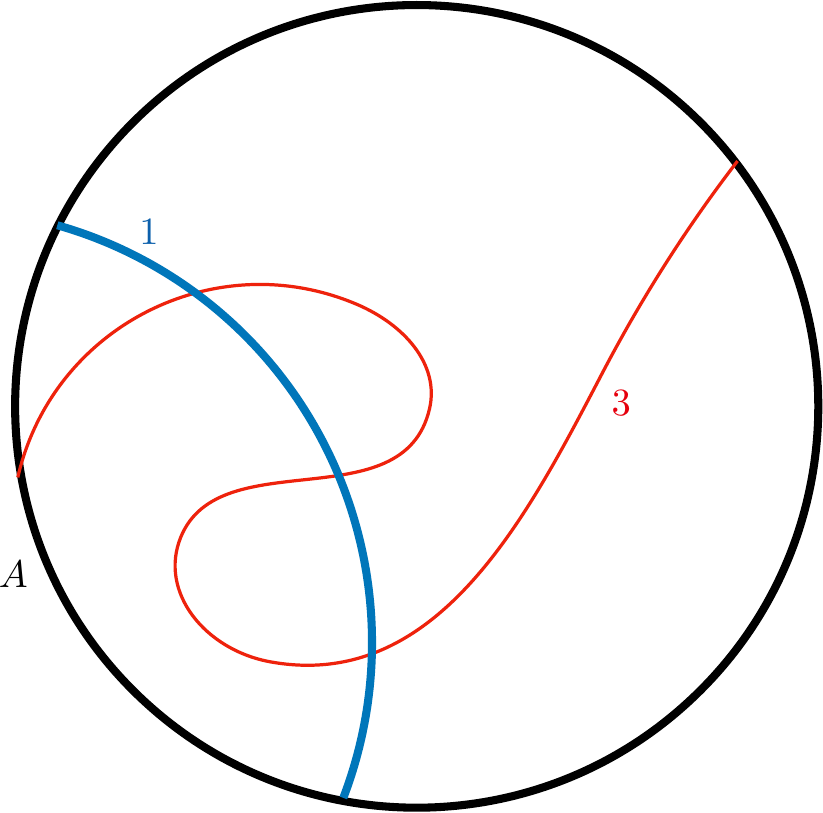}
\caption{\label{fig:CS} The 2-thread shown crosses a total barrier of three. Since the program only requires the thread to cross a total barrier of one, complimentary slackness (CS) implies this thread will not contribute to any optimal configuration.}
\end{figure}
An important consequence of the duality is the concept of complimentary slackness (CS) \cite{boyd2004convex}.  Given the Lagrangian of a convex optimization program with objective $f_0$, constraints $f_i$, and Lagrange multipliers $\lambda_i$ for any optimal configuration it is true that
\be
\lambda_{i}^*f_i(x^*)=0
\ee
which implies one of the two constraints $\lambda_i\geq 0$ or $f_{i}(x)\leq 0$ is saturated. For our program this is the condition
\be
\mu^*(P_{0})=0, \quad P_0 \coloneqq \left\{p \in P:  \int_{\Sigma}d\nu^*(x)\Delta(x,p)> b\right\}
\ee
where $b$ is the required barrier coming from the optimization program (for this program $b=1$). This means for an optimal thread configuration all threads which contribute must cross a total barrier of \emph{exactly} $b$ (see Figure \ref{fig:CS}). Since this will be used throughout the paper to help distinguish the classes of threads and later hyperthreads which will contribute to a given program, we state it as the following theorem:

\begin{theorem}[]
\emph{An optimal thread configuration contains only threads which cross exactly the minimum required barrier $b$. That is: Let $\nu^*$ be an optimal barrier configuration. If a thread $p\in P$ crosses a barrier larger than $b$ then by definition it is in $P_{0}$.  For any dual optimal measure $\mu^*$ we have $\mu^*(P_{0})=0$.}
\end{theorem}

\begin{figure}[H]
\begin{tabular}{cc}
\centering
\includegraphics[width=.45\textwidth,page=13]{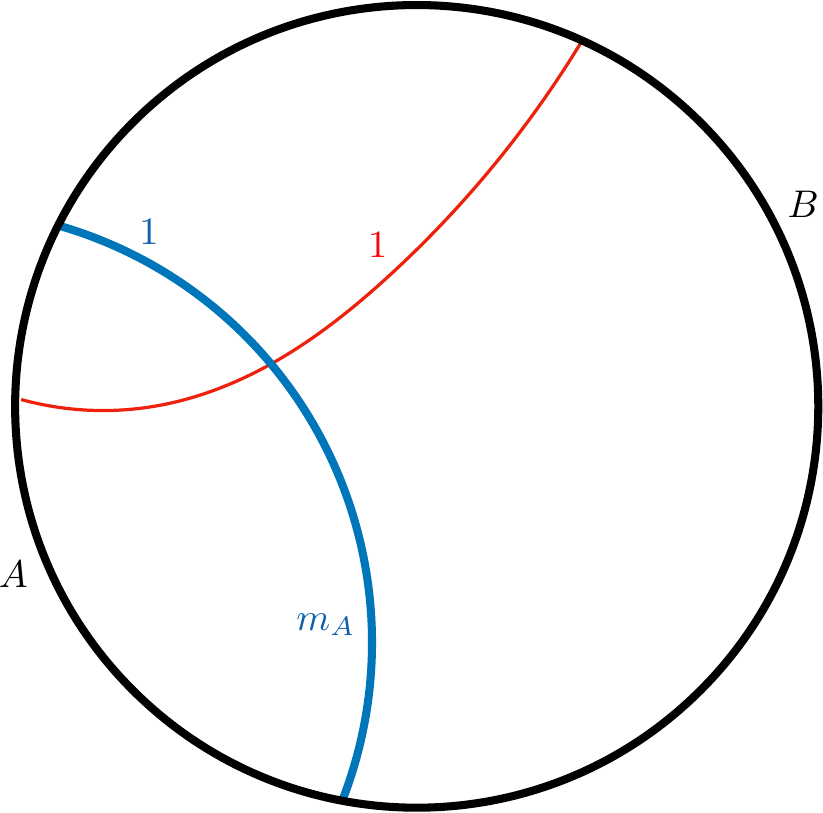}&
\includegraphics[width=.45\textwidth,page=14]{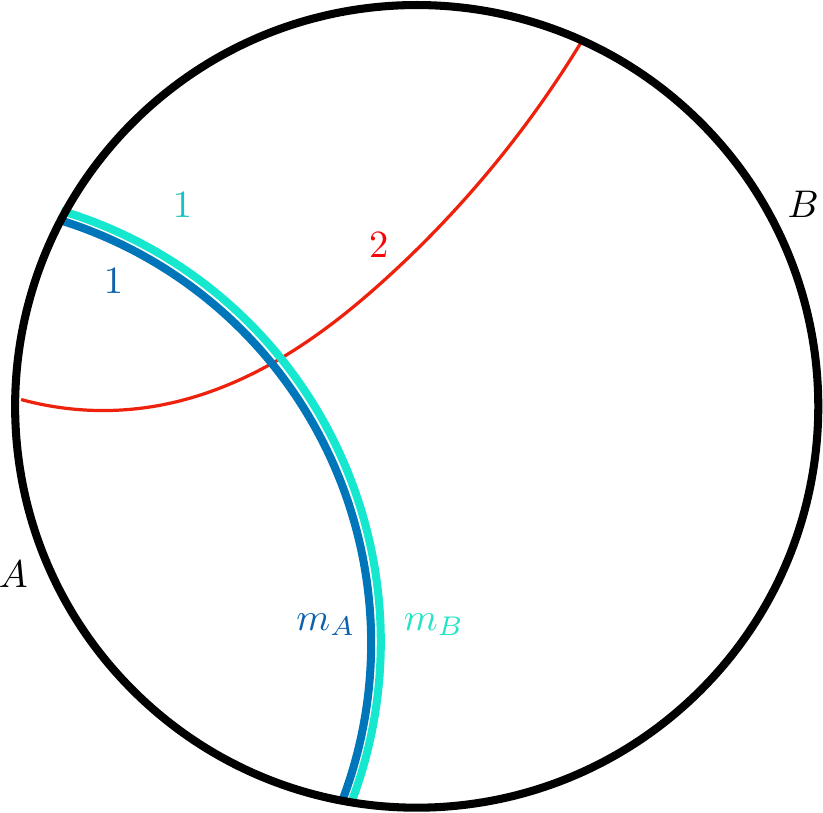}
\end{tabular}
\caption{\label{fig:norm}Two different normalizations which count the number of threads (bell pairs) and thread endpoints (qubits) respectively. For the remainder of this paper we will adopt the latter convention as it more easily generalizes to the multipartite case.}
\end{figure}
As written our optimization programs do not treat the regions $A$ and $B$ the same. This is possible because together $A$ and $B$ comprise the entire boundary so that purity demands $S(A)=S(B)$. As we proceed to more boundary regions this asymmetry in description is no longer useful and it becomes considerably easier to write the programs symmetrically with respect to the boundary regions. This change amounts to a renormalization where instead of counting threads we instead count thread endpoints. When we generalize to hyperthreads this becomes even more important as hyperthreads will connect to the boundary in more than two places. As such if one were to simply count hyperthreads it would not be possible to identify their increased contribution to the objective. Taking this into account we can rewrite \eqref{eq:bimax} and \eqref{eq:bimin} as
\be\label{eq:binorm}
\begin{split}
S(A)+S(B)&=\max 2\mu(P)\text{ s.t. } \forall x\in \Sigma,\; \rho(x) \leq 1\\
&=\min \nu(\Sigma)\text{ s.t. } \forall p\in P,\; \int_{\Sigma}d\nu(x)\Delta(x,p) \geq 2.
\end{split}
\ee
The objective is doubled, reflecting the fact that each thread connects two regions ($A$ and $B$); see Figure \ref{fig:norm}. As a result, the dualization of the minimization program must be modified so that we now require each thread to cross a total barrier of at least 2. This is such that the threads of an optimal configuration saturate both $m_A$ and $m_B$ (which in this case because of purity are the same). We define the surfaces on which a maximal configuration of bit threads saturate to be $m_2$. In this case we have
\be
m_2 = m_A \cup m_B
\ee
so that
\be
2\mu^*=\nu^*=\area(m_2)=S(A)+S(B).
\ee
This can be easily generalized to the case of any number of boundary regions. Let $\mathcal{A}$ be a partition of $\partial\Sigma$ into $n$ non-overlapping boundary regions $A_1,\cdots,A_n$. We define
\be
m_2 \coloneqq \cup_i^nm_{A_i}.
\ee
The program \eqref{eq:binorm} remains unchanged, however the set of bit thread $P$ now includes threads connecting any pair of boundary regions. In order for the constraints to be satisfied the minimal barrier configuration must include a surface which separates each boundary region from the others. As a result the optimal value is the sum of the single party entropies
\be\label{eq:2max}
2\mu^*=\nu^*=\area(m_2)=\sum_i^nS(A_i).
\ee
This result, which is simple to prove in the context of measure theory, is considerably more difficult to show using flows \cite{Cui_2019}.

\section{Hyperthreads}\label{sec:htc}

\begin{figure}[H]
\centering
\includegraphics[width=.5\textwidth,page=1]{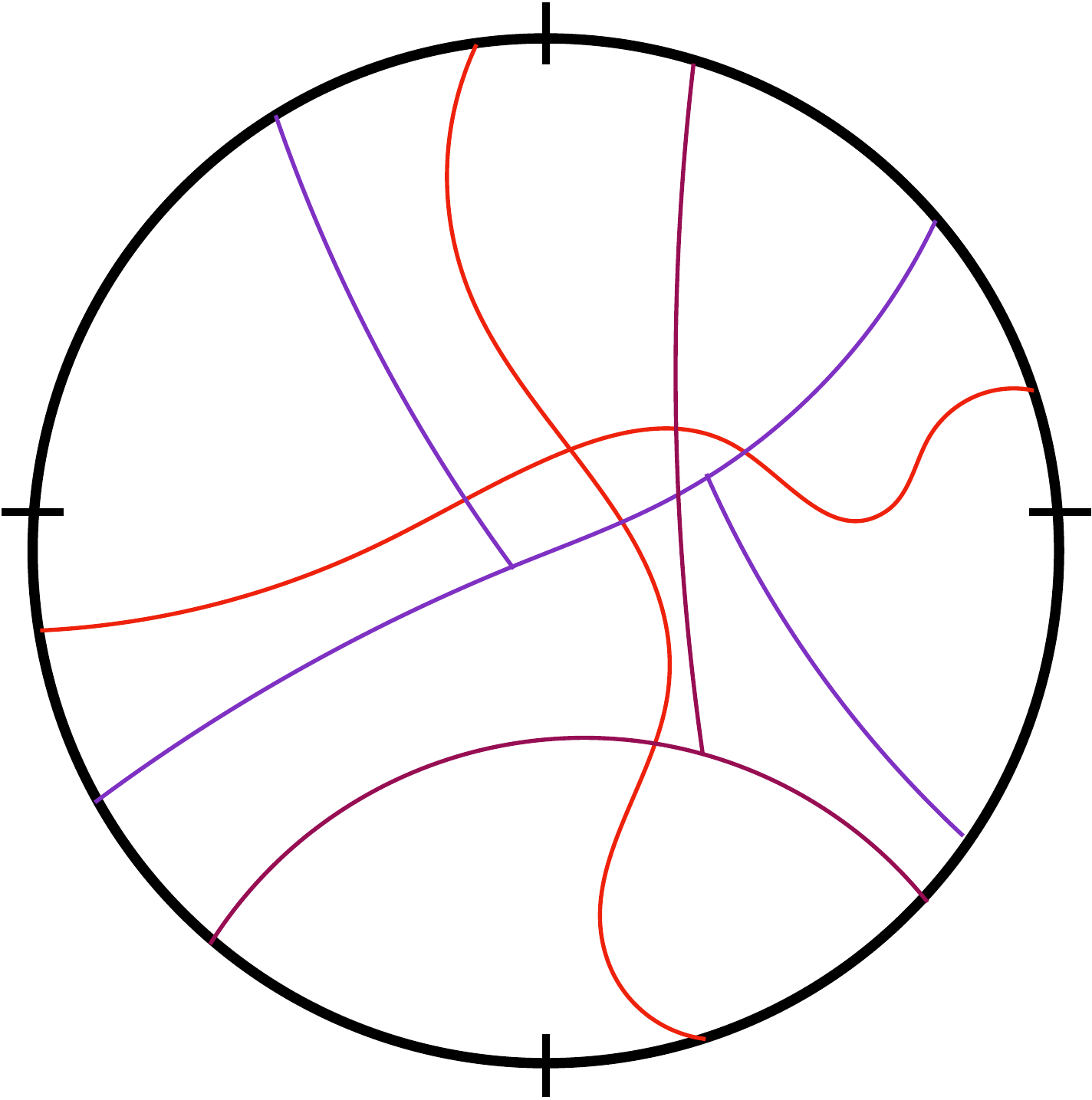}
\caption{\label{fig:HT} Three examples of hyperthreads in $H$ consisting of a single 3-thread and two 4-threads. Note the 4-threads can have either 1 or 2 branching points.}
\end{figure}

Our main goal will be to generalize bit threads in a way suitable for discussing multipartite entanglement. Taking inspiration from hypergraph states \cite{2020ScPP....9...67B,bao2020bit} we define a $k$-hyperthread or ``$k$-thread" $h$ as a connected codimension $1$ subset of $\Sigma$ which forms an embedded tree with between  $1$ and $k-1$ branch points such that the cardinality $\#(h\cap \partial \Sigma)=k$ (see Figure \ref{fig:HT}). Given a partition $\mathcal{A}$ of $\partial \Sigma$  we define the set of all hyperthreads whose endpoints connect to different boundary regions of $\mathcal{A}$ to be $H_\mathcal{A}$.\footnote{As we did for 2-threads, we can always work with the set of all hyperthreads $H_{all}$, which is partition independent, with the additional constraint $\mu(H_{all}\setminus H_{\mathcal{A}})=0$. In what follows we will work with $H_\mathcal{A}$, but omit the subscript and simply call this set $H$ noting that it depends on the boundary partition.}.

It will often be useful for us to talk about hyperthreads with a particular number of boundary endpoints or which connect a specific choice of boundary regions. We define:
\be
H= \cup_{k=2}^n H_k
\ee
where $H_k$ is the set of all $k$-threads and $n=\#(\mathcal{A})$. Each of these can further be decomposed into sets of hyperthreads $H_{A_i:...:A_k}$ which specify the exact regions the hyperthreads connect
\be
\begin{split}
    H_2&=H_{A_1:A_2}\cup...\cup H_{A_{n-1}:A_n}\\
    H_3 &= H_{A_1:A_2:A_3}\cup...\cup H_{A_{n-2}:A_{n-1}:A_{n}}\\
     &\;\;\vdots  \\
    H_n &= H_{A_1:...:A_n}.
\end{split}
\ee
The union for $H_k$ is taken over the $\prescript{}{n}{C}_i$ combinations of elements of $\mathcal{A}$. From this we have $H_2=P$, that is the 2-threads we considered before are a part of $H$. These can also be arranged into sets of hyperthreads which connect to a specific boundary region
\be
H_{A_i} \coloneqq \{h \in H \text{ s.t. } \partial h \cap A_i,A_i^c\neq \emptyset\}.
\ee
For example suppose we have three boundary regions which partition the entire boundary $\partial \Sigma$: $\mathcal{A}=\{A,B,C\}$ then
\be
\begin{split}
   H_2 &= H_{A:B}\cup H_{B:C} \cup H_{A:C} \\
   H_3 &= H_{A:B:C} \\
   H_A &=  H_{A:B} \cup H_{A:C} \cup H_{A:B:C}\\
   H_B &=  H_{A:B} \cup H_{B:C} \cup H_{A:B:C}\\
   H_{AB} &= H_{B:C} \cup H_{A:C} \cup H_{A:B:C} \\
   H_{ABC} &= \emptyset.
\end{split}
\ee
where the others are related by purity (e.g. $H_{AB}=H_{C}$). Note that for example $H_{AB}$ does \emph{not} include $H_{A:B}$ as these hyperthreads do not connect $AB$ to \emph{another} region.

As we proceed to define optimization programs over hyperthreads and their duals, there are several important quantities we will wish to distinguish. Suppose we have an optimization program $O$ whose optimal configuration for the dual and primal is a collection of $k$-threads which saturate on a surface $s$ with barrier $b$. In general, as we will see, a $k$-thread will typically contribute $k$ to the objective while crossing the barrier $\alpha$ times. As such the area of the surface $s$ is
\be
\area(s),
\ee
while number of hyperthreads in the configuration will be given by this area divided by the number of times $\alpha$ that a hyperthread crosses
\be
\frac{1}{\alpha} \area(s).
\ee
The optimal value of the objective is given the number of hyperthreads times their contribution. Because of the duality this will be the same as the total barrier
\be
opt(O)=\frac{k}{\alpha} \area(s)=b\area(s).
\ee

The plan for the remainder of this section is as follows: We will describe two different dual optimization programs. The first of these is over the entire set $H$ including 2-threads. We will show that the optimal configuration consists primarily of 2-threads with a most a finite number of $k$-threads with $k\geq3$. For the second, we will consider a program only over the space of $k$-threads, $H_k$, for a fixed value of $k$. Here we will see the behavior of the optimal configuration can be quite different from the previous case. Since the resulting configuration contains only $k$-threads, this will motivate us to define a potential measure of multipartite entanglement which we call the hyperthread partition $HP(\mathcal{A})_k$.

The dual programs we define are general enough such that they hold for any choice of spacetime. However, our goal here is not to exhaustive, but rather to provide concrete examples of the behaviors that hyperthreads can exhibit. To this end we will consider several distinct geometries: vacuum AdS$_3$, AdS$_3$ black holes, and multiboundary wormholes. The behavior of the programs in these cases is drastically different which stems from the existence of the uv cutoff in AdS$_3$. Similar analysis to that provided can be applied to other geometries or dual states of a holographic CFT which one might be interested in.

\subsection{Maximal hyperthread configurations on $H$}
Given the definitions above we can extend the program \eqref{eq:bimax} to include hyperthreads. It is important to recognize that we are not actually interested in the number of hyperthreads, but rather the total number of connections the threads have with the boundary. Thus, we should consider the objective
\be
\sum_{i=1}^n\int_{H_{A_i}}d \mu(h) = \sum_{k=2}^nk\int_{H_k}d\mu(h)
\ee
where we have used the definition of $H_{A_i}$ and $H_{k}$. This resummation makes sense as each $k$-thread should contribute $k$ times to the objective. Using this we define the maximization program
\be\label{pgm:weightedH}
\max \sum_{k=2}^nk\int_{H_k}d\mu(h) \text{ s.t. } \forall x\in \Sigma,\; \int_{H}d\mu(h)\Delta(x,h)\leq 1,
\ee
which maximizes the total number of hyperthread endpoints subject to the density constraint. In the case $H=H_2$ (we only consider 2-threads) this reduces to \eqref{eq:bimax}. The dualization of the program \eqref{pgm:weightedH} proceeds as before

\be
\begin{split}
L(d\mu,d\nu)&= \sum_{k=2}^nk\int_{H_k}d\mu(h)-\int_{\Sigma}d\nu(x)\left(\int_{H}\left(d\mu(h)\Delta(x,h)\right)- 1\right)\\
&=\sum_{k=2}^n\int_{H_k}d\mu(h) \left(k-\int_{\Sigma}d\nu(x)\Delta(x,h)\right)+\int_{\Sigma}d\nu(x)
\end{split}
\ee
from which we have the dual minimization program
\be\label{pgm:minH}
\min \nu(x)\text{ s.t. } \forall_k^n:\;\forall h\in H_k,\; \int_{\Sigma}d\nu(x)\Delta(x,h) \geq k.
\ee
This asks for a barrier configuration such that each $k$-thread crosses a total barrier of at least $k$. The solution to this program is given by $\area(m_2)$ which can be seen by noting that \eqref{pgm:minH} is the natural generalization of \eqref{eq:binorm} with additional constraints imposed on the hyperthreads with $k>2$. As a result these additional constraints can only make the optimal objective larger. Thus,
\be
\area(m_2)\leq \nu^*
\ee
\begin{figure}[H]
\centering
\includegraphics[width=.5\textwidth,page=3]{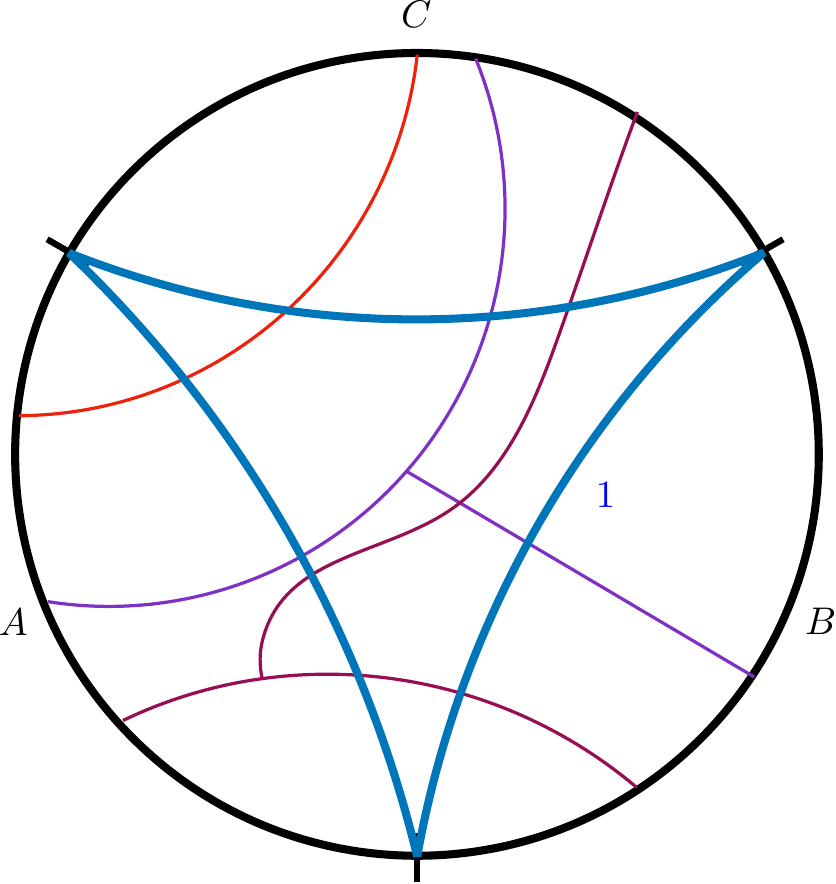}
\caption{\label{fig:m2} An example for three regions in AdS$_3$. All 2-threads will cross a barrier of 2. There are two classes of 3-threads the first splits in one of the homology regions $r(A_i)$ crosses a total barrier of 4 and will not contribute to the optimal configuration. The rest split in the interior region $\Sigma\setminus\cup_ir(A_i)$ and cross a total barrier of 3; these can contribute.}
\end{figure}
Finally, we see that the barrier configuration on $m_2$ in fact satisfies the additional constraints on the hyperthreads. That is any $k$-thread will cross $m_2$ at least $k$ times. The $k$-threads which cross the minimal barrier are those which split in the interior region $\Sigma\setminus\cup_ir(A_i)$ (see Figure \ref{fig:m2}). That is, we have shown that the dual program \eqref{pgm:weightedH} and \eqref{pgm:minH} has the optimal value
\be
\sum_{k=2}^nk\mu^*(H_k)=\nu^*=\area(m_2)=\sum_i^nS(A_i).
\ee

\paragraph{Contributions from hyperthreads}
Given the program \eqref{pgm:weightedH} and \eqref{pgm:minH} it is natural to ask how much $k$-threads can contribute. It is already known that $m_2$ can be saturated using only 2-threads (see \eqref{eq:2max}) so what is desired is to steer the program towards other maximal configurations which contain $k$-threads. To do so we modify the program by adding an additional term to the objective which attempts to maximize over the $k$-threads. By giving this term an infinitesimal weight $\epsilon$ we guarantee that the optimal configuration has as many $k$-threads as possible without changing the original objective. For simplicity we assume that we have chosen a partition of the boundary into $n$ regions and consider the contribution of $n$-threads to the objective \footnote{The same procedure works similarly if we instead perturb $H_k$ with $2<k\leq n$.}
\be\label{pgm:weightedHP}
\max \sum_{k=2}^{n-1}k\int_{H_k}d\mu(h)+(n+\epsilon)\int_{H_n}d\mu(h) \text{ s.t. } \forall x\in \Sigma,\; \int_{H}d\mu(h)\Delta(x,h)\leq 1.
\ee
The only change after dualizing is an increase in the required barrier for the $n$-threads by $\epsilon$
\be\label{pgm:minweightedHP}
\min \nu(x)\text{ s.t. } \forall_k^{n-1}:\;\forall h\in H_k,\; \int_{\Sigma}d\nu(x)\Delta(x,h) \geq k,\;\forall h\in H_n,\; \int_{\Sigma}d\nu(x)\Delta(x,h) \geq n+\epsilon.
\ee
\begin{figure}[H]
\begin{tabular}{cc}
\centering
\includegraphics[width=.45\textwidth,page=12]{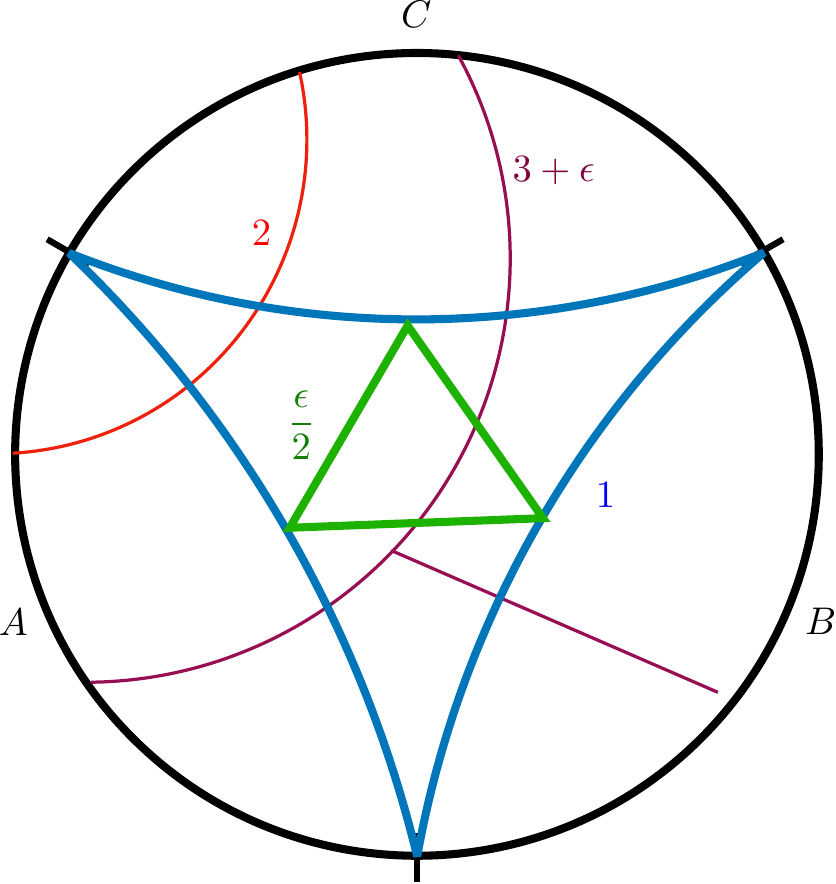}&
\includegraphics[width=.45\textwidth,page=22]{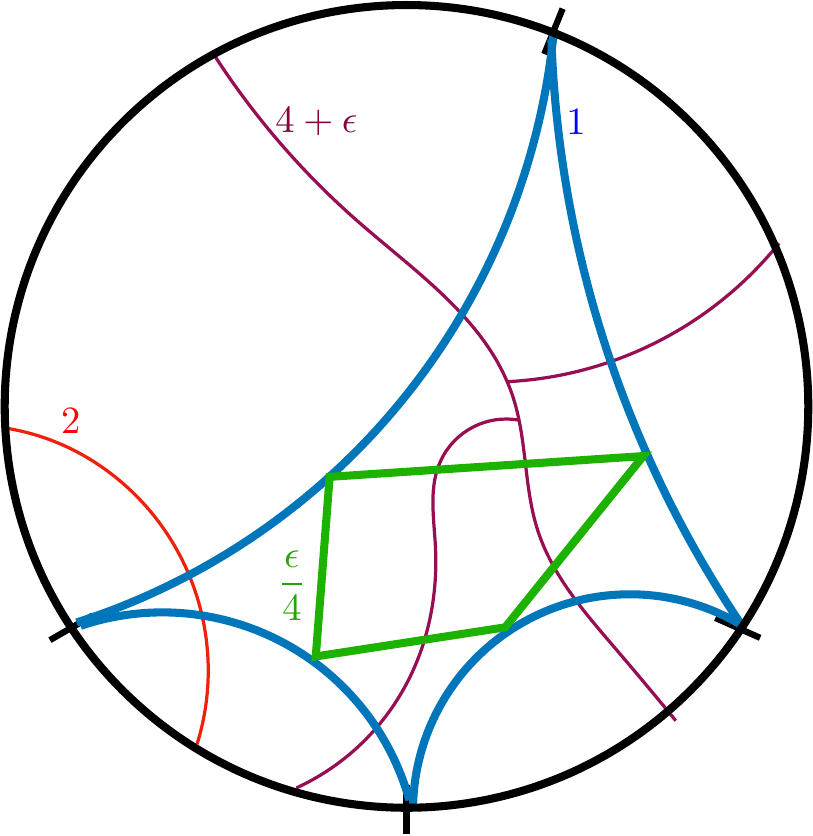}
\end{tabular}
\caption{\label{fig:perturbedEE} L: An example for AdS$_3$ with three boundary regions. We look for a maximal hyperthread configuration with the added condition that we want as many 3-threads as possible. The minimal surface configuration to this perturbed problem is given by $m_2$ with the 3-threads saturating the green triangle, $t_3$, which is finite. Complimentary slackness guarantees that any 2-threads which cross any piece of $t_3$ will not contribute. This essentially combs the 2-threads away from the center so that the 3-threads will reside there. Similarly, any 3-threads which would split inside $t_3$ will not contribute. R: Another example for four unequal boundary regions. The maximum number of 4-threads is given by $\frac{1}{4}\area(t_4)$.}
\end{figure}
From this we have the optimal solution
\be\label{maxhp}
\sum_{k=2}^{n-1}k\mu^*(H_k)+(n+\epsilon)\mu^*(H_n)=\nu^*=\area(m_2)+\frac{\epsilon}{\alpha} \area(t_n)
\ee
where $t_n$ is defined to be the surface where the additional barrier proportional to $\epsilon$ was placed in the optimal configuration. The factor $\alpha$ accounts for the number of times an $n$-thread crosses $t_n$. CS guarantees that \emph{only} $n$-threads cross this surface as any 2-thread which crossed it would not contribute to the optimal configuration because of the additional barrier. Thus the area of $t_n$ is proportional to the maximum number of $n$-threads which can contribute to the entanglement entropy (see Figure \ref{fig:perturbedEE}).

As $\epsilon$ goes to zero we recover the optimal value  $\area(m_2)$ with a configuration of hyperthreads consisting of:
\begin{equation*}
\begin{split}
    \frac{1}{\alpha}\area(t_n) \text{ $n$-threads each contributing $n$}\\
    \frac{1}{2}\left(\area(m_2)-\frac{n}{\alpha}\area(t_n)\right) \text{ 2-threads each contributing 2.}
\end{split}
\end{equation*}
From this it is apparent that the contribution of hyperthreads to the calculation of the entanglement entropy  in AdS$_3$ is generically finite. We conjecture that this holds more widely. This can be verified by the application of the dual programs \eqref{pgm:weightedHP}, \eqref{pgm:minweightedHP} to other spacetimes of interest.




\subsection{Maximal hyperthread configurations on $H_k$}
Next, we will consider programs which only contain $k$-threads for a single fixed value of $k$. This is done by instead optimizing the program over the set $H_k$ so that we get
\be\label{pgm:hk}
\begin{split}
HP_k(\mathcal{A})&=\max\; k\mu(H_k)\text{ s.t. } \forall x\in \Sigma,\; \int_{H}d\mu(h)\Delta(x,h)\leq 1\\
&=\min \nu(x)\text{ s.t. } \forall h\in H_k,\; \int_{\Sigma}d\nu(x)\Delta(x,h) \geq k.
\end{split}
\ee
The program requires a minimal barrier configuration such that all $k$-threads cross a barrier of at least $k$. Though very similar we will see that the lack of 2-threads leads to very different behavior. We have defined the optimal value of this program to be the hyperthread partition $HP(\mathcal{A})_k$. To aid in the description of the optimal configurations of this program we will consider examples in both vacuum AdS$_3$ and multiboundary wormhole geometries constructed by identifying geodesics. 

\subsubsection{Vacuum AdS$_3$}
We first consider the dual programs \eqref{pgm:hk} in vacuum AdS$_3$ and consider a partition of the boundary into three equal sized regions. In this case we are interested in maximizing over the set of 3-threads $H_3$.
\begin{figure}[H]
\begin{tabular}{cc}
\centering
\includegraphics[width=.45\textwidth,page=17]{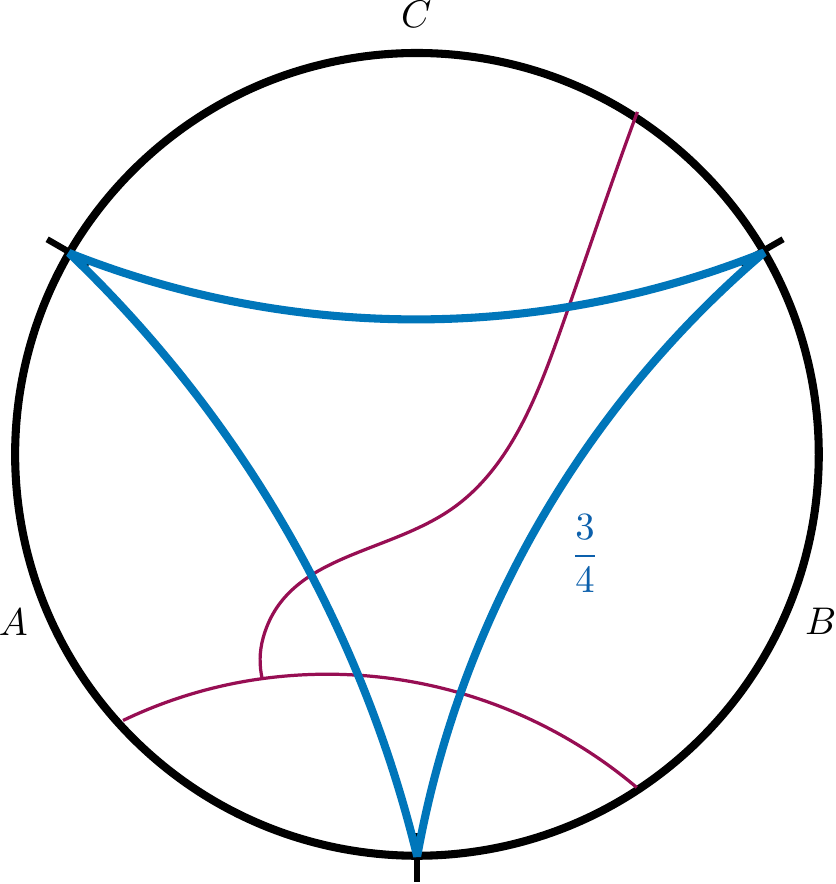}&\includegraphics[width=.45\textwidth,page=18]{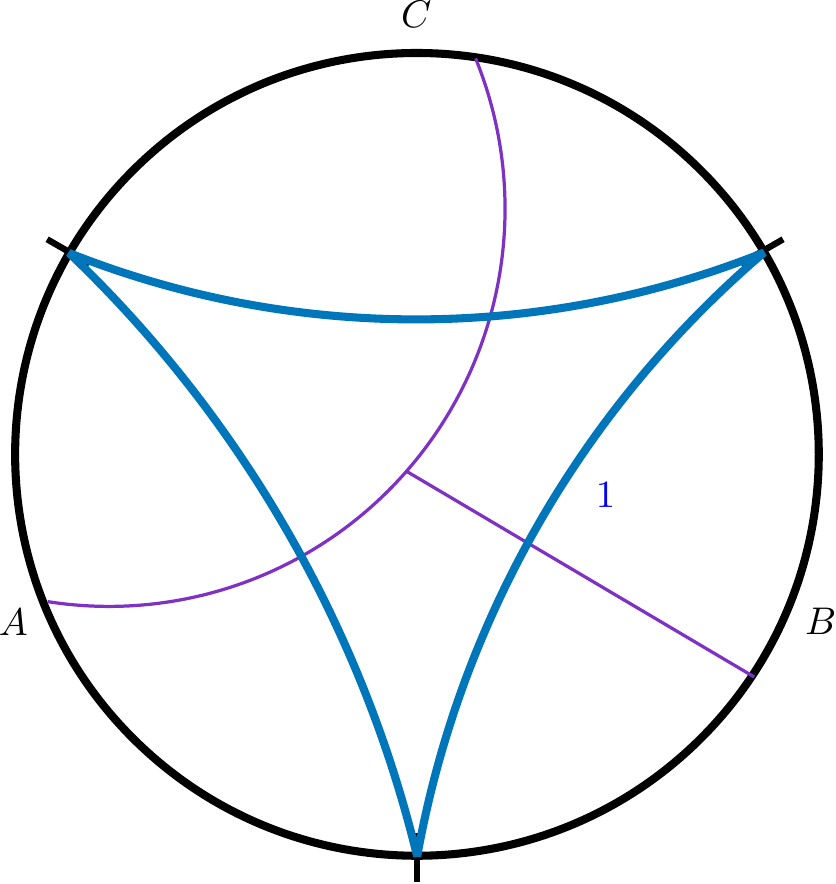}\\
\includegraphics[width=.45\textwidth,page=19]{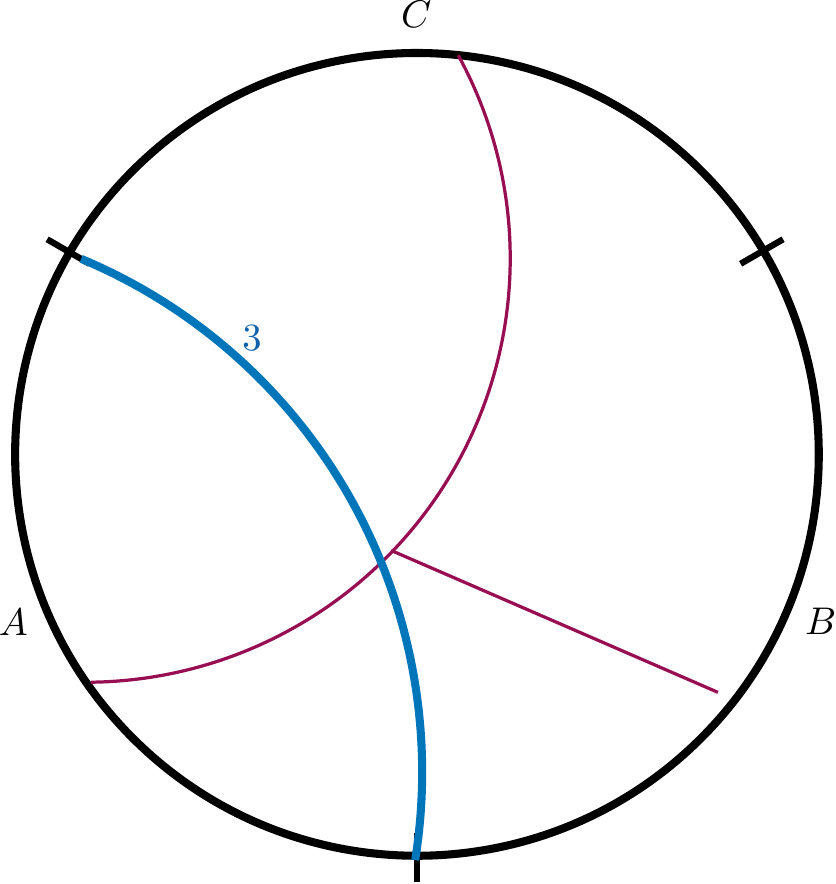}&\includegraphics[width=.45\textwidth,page=5]{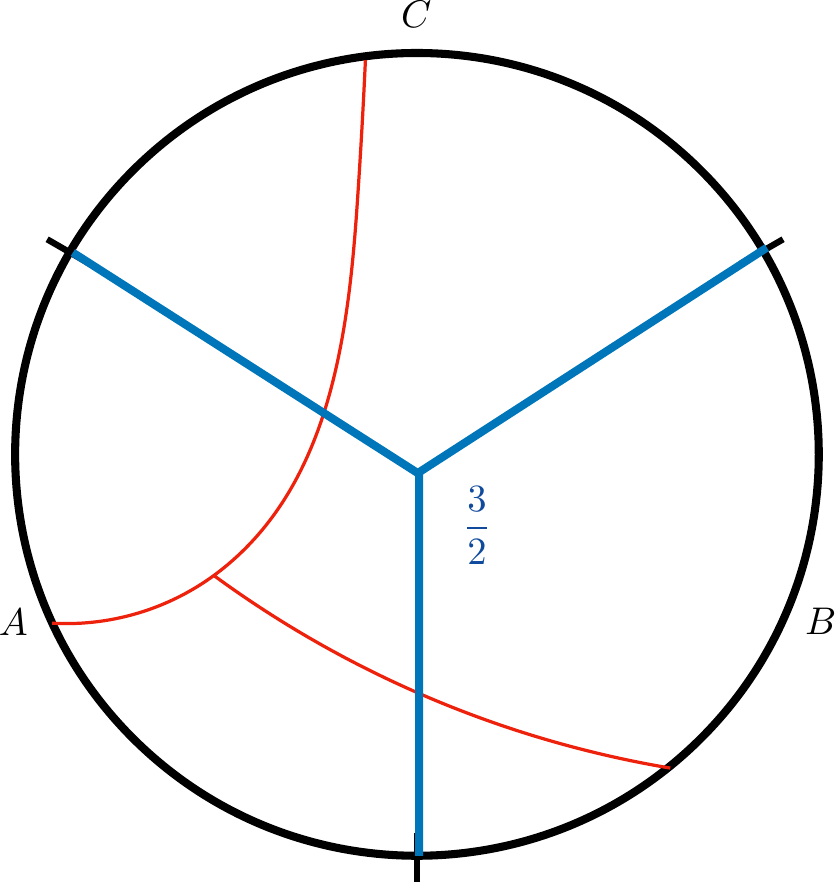}
\end{tabular}
\caption{\label{fig:m3} A number of possible barrier configurations for the program \eqref{pgm:hk} for three regions of AdS$_3$. Shown in each is an example of a 3-thread which could contribute to the objective. TL: The configuration is not feasible as a 3-thread splitting in the interior region would cross a barrier of $\frac{9}{4}<3$. TR: The proposed configuration is given by $m_2$ however \eqref{maxhp} proved $3$-threads cannot saturate this surface. BL: Placing the barrier on the smallest RT surface is feasible, but is large due to the contribution from the uv cutoff at the boundary. BR: The true minimal surface $m_3$. The introduction of a bulk intersection point substantially reduces contribution coming from the boundary.}
\end{figure}
\noindent Any choice of barrier which creates an interior and exterior region (see the top of Figure \ref{fig:m3}) will create two classes of 3-threads depending on which bulk region the 3-thread splits. This excludes a large number of 3-threads from being able to contribute to the objective. As such the minimal barrier configuration will be such that there is no interior bulk region while remaining symmetric with respect to the boundary regions. The conclusion of this line of reasoning is that the minimal barrier configuration must split the entire bulk into three regions, one of which is homologous to each boundary. In other words, the constraints on the hyperthreads force us to consider a very particular class of surfaces. The minimal surface with this property is the ``Mercedes diagram" $m_3$ (see the bottom of Figure \ref{fig:m3}).\footnote{The surface $m_3$ can also be related to a dual program over three vector fields with the same nonzero divergence (see Appendix C of \cite{Harper_2019}). Each vector field can be viewed as a collection of one of the three segments of the hyperthreads with the nonzero divergence acting as the vertex which connects them. In this paper, flow programs over multiple vector fields were also defined for the multipartite entanglement wedge cross section (MEWC). It seems plausible that these may also be able to be reinterpreted as hyperthreads. This is consistent with the current understanding that the MEWC (including the bipartite case) requires multipartite entanglement \cite{2020JHEP...04..208A,2021arXiv210700009H}. We leave further exploration to future work.} More generally we define $m_k$ to be the minimal surface on which a maximal configuration of $k$-threads saturates. For this example the maximum number of 3-threads is given by $\frac{1}{2}\area(m_3)$.
 
 

\begin{figure}[H]
\begin{tabular}{cc}
\centering
\includegraphics[width=.45\textwidth,page=21]{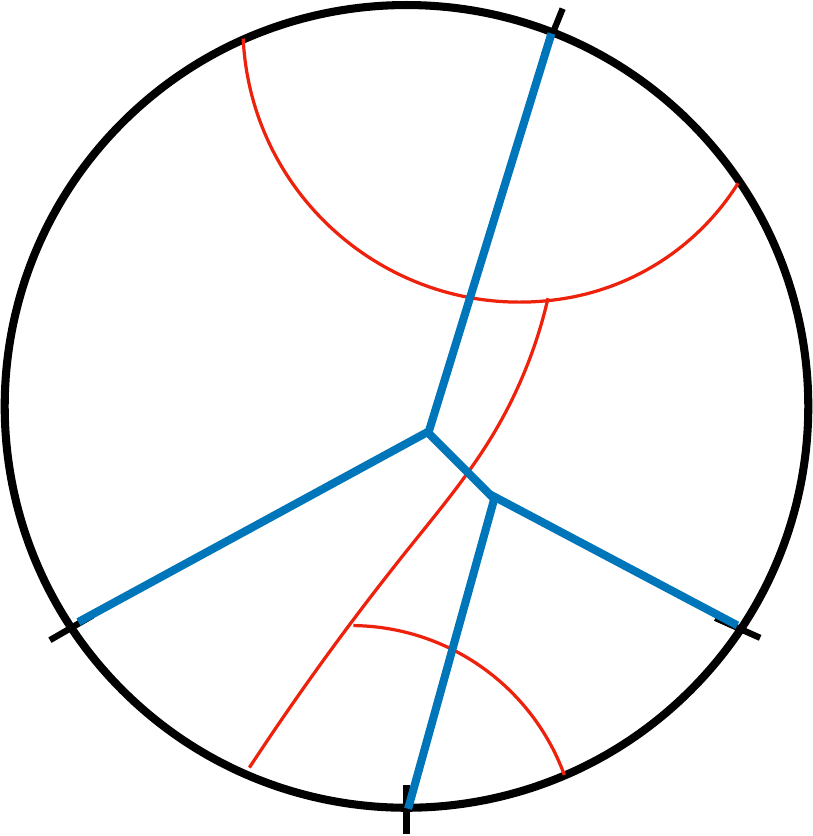}&
\includegraphics[width=.45\textwidth,page=20]{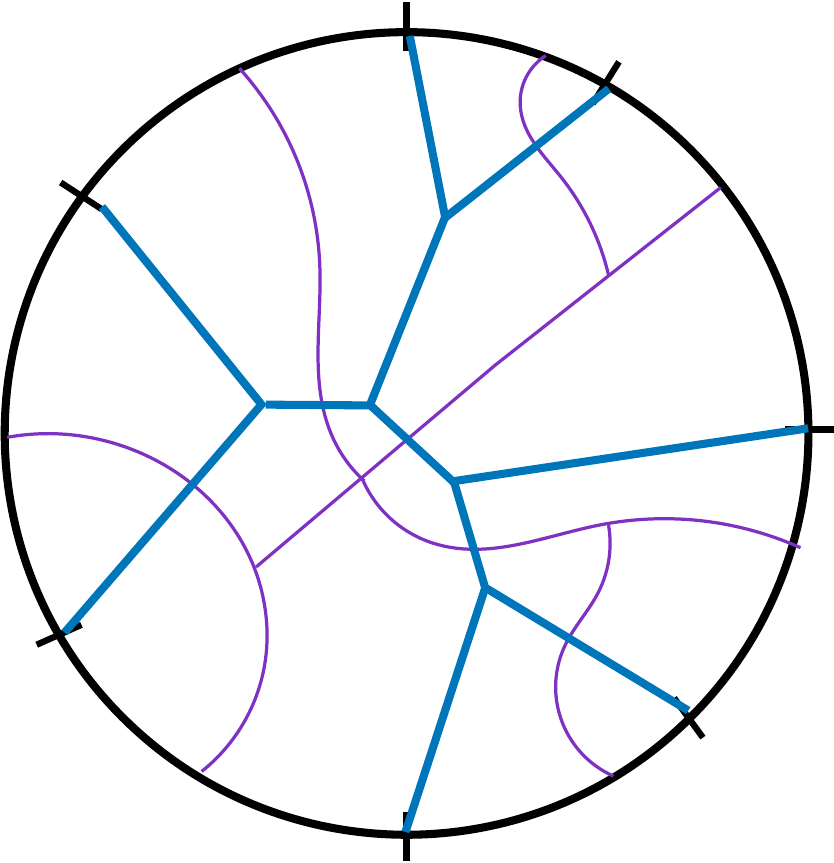}
\end{tabular}
\caption{\label{fig:m4} Potentially minimal boundary configurations $m_4$ for 4 boundary regions and $m_7$ for 7 boundary regions. Both surfaces are constructed using the double bubble solution on $\mathbb{H}_2$. Also shown in each is an example of a hyperthread which could contribute.}
\end{figure}
 We now move on to more general boundary partitions. We first note that the inclusion of bulk intersections for a barrier configuration reduces the barrier that must be placed on the boundary. Because of the uv cutoff such contributions are infinitely large so that the minimization program avoids these at all costs. This means that the minimal configuration will try to include as many bulk intersections as possible. So that all the hyperthreads will cross the barrier configuration the same number of times each boundary region is necessarily segregated from the others so that there is a single bulk region associated to each boundary. Thus, the problem reduces to constructing a bulk partition. The minimization of such a partition is a long studied problem in geometric measure theory with a conjectured solution, the ``double bubble":
 \begin{conjecture}[Double bubble]
 The least area connected partition of a manifold $M$ into any number of regions consists of the union of minimal surfaces such that their intersections in $M$ are trivalent and equiangular.
 \end{conjecture}
 \noindent For the purposes of this paper we note that this is proven to hold for the hyperbolic disk $\mathbb{H}^2$ and conjectured to hold for more general manifolds in particular $\mathbb{H}^n$ which is the case of interest for constant time slices of AdS$_{n+1}$ \cite{10.2307/3062123,cotton}. With this in mind it is now straightforward to construct minimal barrier configurations for any number of boundary regions (see Figure \ref{fig:m4}). In fact the double bubble solution is such that it satisfies the necessary constraints regardless of which $H_k$ we optimize over. That is for AdS$_3$: $m_3=\cdots=m_n$. Where the difference in the optimal value is due to the contribution, $k$, and number of crossings, $\alpha$, of a $k$-thread:
 \be
 HP(\mathcal{A})_{k}=\frac{k}{\alpha}\area(m_k)
 \ee
 where here for AdS$_3$ this can be reduced to
 \be
 HP(\mathcal{A})_{k}=\frac{k}{k-1}\area(m_n).
 \ee
 
\subsubsection{Black hole states}

Next, we will consider a single sided black hole in AdS$_3$ which is dual to a mixed state of the boundary CFT. To be concrete we consider the division of the boundary into two regions $\mathcal{A}=\{A,B\}$ where we choose $A$ to be the smaller of the two. In order for the threads to account for the thermal contribution which is given by the area the black hole horizon $\sigma$ it is necessary for the threads to be able to end there. This can be justified by considering a purification of the geometry to the thermofield double. In this picture the threads which end on the horizon can be continued to the other side to the new boundary $O$. Thus, even with two boundary regions in the original geometry when we consider programs of hyperthreads we will choose the space of hyperthreads $H$ to include 2-threads $H_{A:B},H_{A:O},H_{B:O}$ as well as the 3-threads $H_{A:B:O}$. We can now consider the application of the programs \eqref{pgm:minweightedHP} and \eqref{pgm:hk} to this state. 

\begin{figure}[H]
\centering
\includegraphics[width=.5\textwidth,page=29]{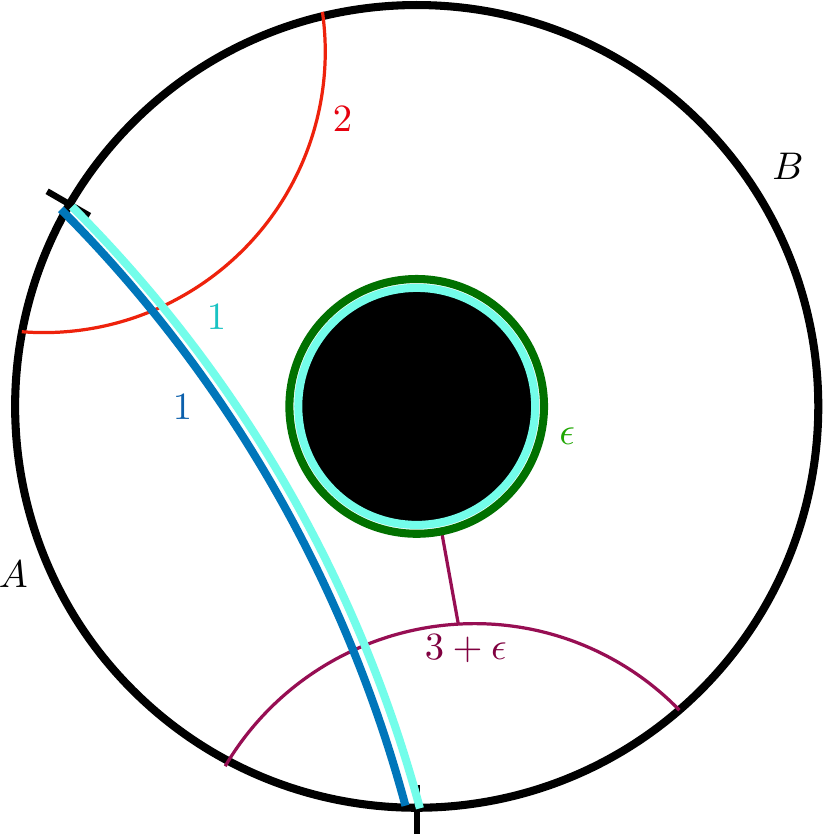}
\caption{\label{fig:BH} The minimal barrier configuration of \eqref{pgm:minweightedHP} for the black hole state is given by $m_A\cup m_B$ which includes the black hole horizon. Since the 3-threads must connect to horizon which is finite this surface acts the limiting factor for their contribution. As such the maximal hyperthread configuration includes a number of 3-threads up to thermal entropy along with additional 2-threads connecting $A$ to $B$. The barrier configuration excludes 2-threads connecting either $A$ or $B$ to the horizon}
\end{figure}

As in the case of pure AdS$_3$ the program \eqref{pgm:minweightedHP} has an optimal value equivalent to the sum of the entanglement entropies which includes the area of the black hole horizon
\be
\sum_{k=2}^nk\mu^*(H_k)=\nu^*=\area(m_2)=S(A)+S(B)=2S(A)+\area(\sigma).
\ee
 Since the horizon if finite (compared to $m_A$ which is uv divergent) it will be the surface which limits the contribution of 3-threads (see figure \ref{fig:BH}). This means that in the optimal hyperthread configuration the number of 3-threads is given precisely by the area of the black hole horizon.

\begin{figure}[H]
\centering
\includegraphics[width=.5\textwidth,page=30]{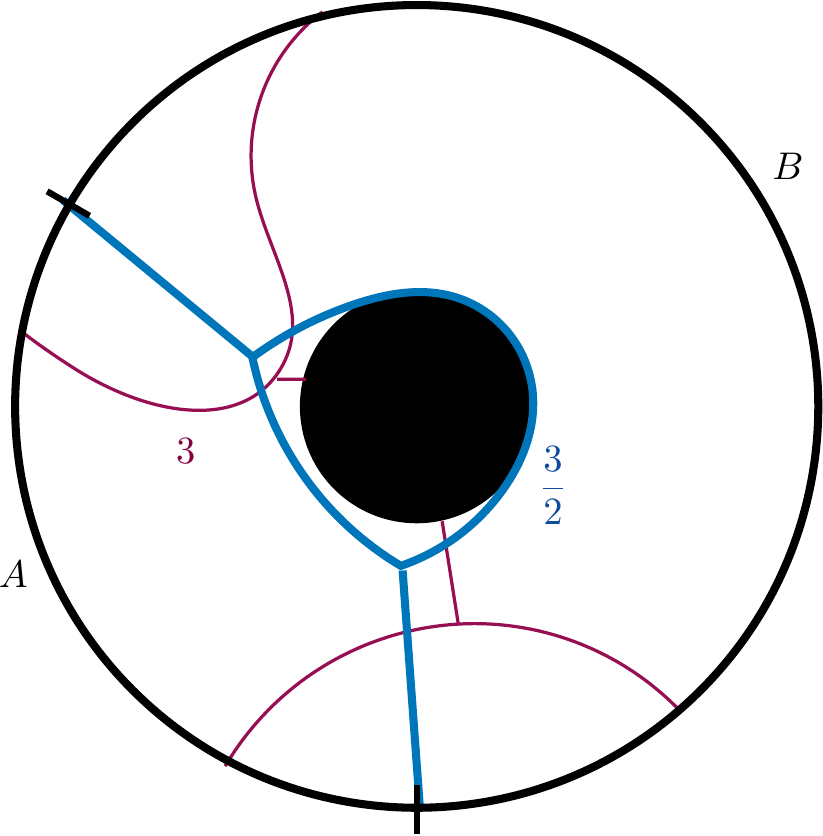}
\caption{\label{fig:BHHP} The minimal barrier configuration for \eqref{pgm:hk} applied to a black hole state with $k=3$. The solution is found using the double bubble conjecture with the black hole horizon (or equivalently the purifying boundary region $O$) acting as the third region. Generically the minimal barrier configuration will deform such that it is not commensurate with the black hole horizon. Shown are examples of potentially contributing 3-threads which connect to the horizon, $A$ and $B$.}
\end{figure}

The analysis of the program \eqref{pgm:hk} proceeds analogously to vacuum AdS$_3$. Considering $A,B,\sigma$ to constitute the full boundary of the manifold a minimal barrier configuration can be constructed with the aid of the double bubble solution (see figure \ref{fig:BHHP}). In general the resulting barrier will \emph{not} lie on the horizon as the various competing surfaces will deform to find the global minimum.

\subsubsection{Multiboundary wormholes}
As our final example, we consider a particular class of geometries:  multiboundary wormholes \cite{Krasnov:2000zq,Krasnov:2003ye,Skenderis:2009ju,Balasubramanian:2014hda,Maxfield_2015}. These spacetimes are formed by choosing a collection of boundary regions of ${\rm AdS}_3$ and identifying various internal geodesics that bound the resulting homology region. As a result of this procedure the geometry no longer has a uv cutoff and the entanglement entropies are finite. Compared to vacuum AdS$_3$ the lack of a uv cutoff indicates there is no longer a benefit for minimal boundary configurations to contain bulk intersection points so that the behavior of \eqref{pgm:hk} is decidedly different.
\begin{figure}[H]
\centering
\includegraphics[width=.5\textwidth,page=15]{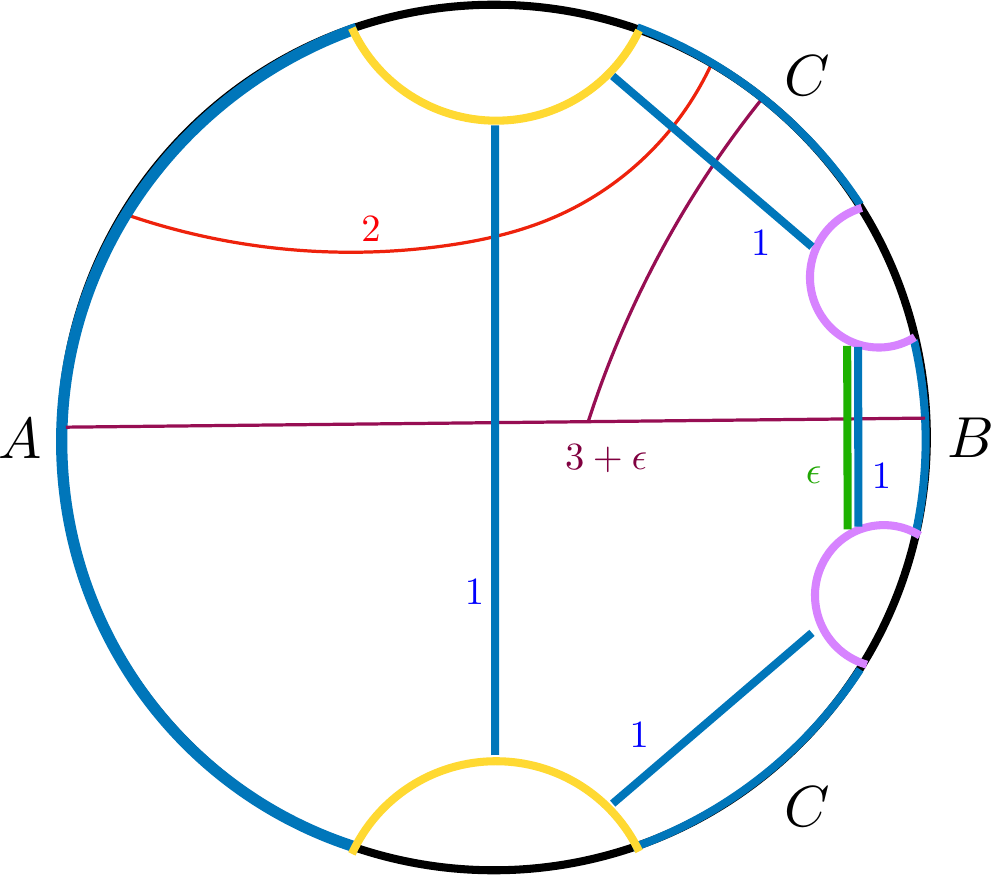}
\caption{\label{fig:MBWH} A 3-boundary wormhole constructed by identifying geodesics (the yellow and purple pairs) of a constant time slice of AdS$_3$. Using the methods described above we find that the maximum number of 3-threads is given by the smallest entanglement entropy of the three boundary regions. Here we take it to be $S(B)$. In fact a configuration of $2$ and $3$-threads which saturates $m_2$ can be found such that we have the maximum possible 3-threads i.e. $t_3=m_3=m_B$. Unlike in the case of vacuum AdS, since the entanglement entropy is finite the contribution of 2-threads and $k$-threads can be chosen to be of comparable size.}
\end{figure}

Given a multiboundary wormhole geometry with $n$ boundaries in the fully connected phase, the entanglement entropy of any disjoint union over complete boundaries is given by a union of wormhole throats, the minimal surfaces homologous to them. First considering an optimization over $H_n$ we see that each $n$-thread will cross each wormhole throat so that it connects all $n$ boundaries. The limit to the number of $n$-threads is then given by the smallest of the wormhole throats. As a result, it is in fact possible to construct a configuration of 2-threads along with the maximum number of $n$-threads so that together they saturate on $m_2$ (see Figure \ref{fig:MBWH}). That is, we have
\be
t_n=m_n,\; \area(m_n)=\min_{\mathcal{A}}\area(m_{A_i}).
\ee

In the case where we instead optimize over $H_k$ with $2<k<n$ we can reach a similar conclusion. Now since $k\leq n$ a $k$-thread will only cross $k$ of the wormhole throats. Suppose we pick $k$ of the $n$ boundary regions and first maximize $H_{A_1:\cdots:A_k}$, that is the hyperthreads connecting exactly the boundary regions we chose. The most we can include will be the smallest of these $k$ throats with the others partially saturated. We can repeat this again with another selection of $k$ boundary regions excluding the one which is saturated until one of these $k$ throats saturates. This procedure can be continued with different collections of boundary regions until no more $k$-threads can be placed. Maximizing over all such constructions shows that the bottleneck will always be given by a sum over some subset of the ``small" single party entanglement entropies where the details depend on relative sizes of the $n$ throats. The remaining space can then be filled in with 2-threads such that the configuration saturates on $m_2$. What we have shown is for multiboundary wormholes we can always find a configuration of 2-threads and $k$-threads saturating on $m_2$ such that
\be
t_k=m_k, \; \mathcal{O}(\area(m_k)) \sim \mathcal{O}(S(A_i))
\ee
so that the 2-thread and $k$-thread contributions to the entanglement entropy are of similar size. This indicates for this class of geometries particular sums of entanglement entropies (corresponding to a hyperthread partition) may be useful for diagnosing multipartite entanglement.
 
\section{Discussion}\label{sec:dis}
In this article we have defined hyperthreads and shown that their maximal configurations are equivalent to various classes of minimal surfaces in the bulk. What remains to be done is to connect these surfaces to measures of multipartite entanglement in the boundary CFT. This is a hard problem due to the complex nature of classifying multipartite entanglement. In this section we conclude with some speculative topics assuming this connection to be generically true. Of course, this has yet to be proven as one would need to determine a useful way to characterize the multipartite entanglement of a CFT state, define the associated resource theory and accompanying measures, and explicitly demonstrate an ``RT-like" duality possibly through the use of replica trick methods. This is a large task, but one we hope future efforts will be able to tackle.

 First, we discuss the multipartite distillation of CFT states. Given that a bit thread is equivalent to a distilled bell pair (this relies intimately on the fact that the distillation is bipartite) we assume a suitable generalization that each hyperthread can be thought of as a distilled unit of $k$-party entanglement. If true this implies a number of consequences which follow from the various optimization programs we have defined. 
 
 Next, we describe a possible procedure for generating ``geometries" which allow for large numbers of hyperthreads. This is essentially the continuum equivalent of a hypergraph. Notably, the constructed example does not satisfy MMI which allows for large amounts of genuine multipartite entanglement. 

Finally, we describe a generalization of hyperthreads which allow for threads to contribute negatively to both the objective and density bound. It is shown that these ``negative threads" along with the usual ``positive threads" can be used together to generically circumvent geometric bulk obstacles. It is possible this may be useful for the definition of hyperthread duals for other bulk surfaces of interest. 

\subsection{Multipartite distillation of CFT states}
Throughout this paper we have been careful not to refer to a configuration of hyperthreads as a distillation. Remember that in the case of 2-threads the threads, should be thought of as a particular distillation of the CFT state into bell pairs using local unitaries (LU), a subset of local operators and classical communication (LOCC), asymptotically over many copies of the system
\be
\ket{\psi}\xrightarrow{LU}\ket{2}_{AB}^{\otimes S(A)}
\ee
here we have defined the cat-$i$ state between $i$ different parties
\be
\ket{i} \coloneqq \frac{1}{\sqrt{2}}\left(\ket{0\dots0}+\ket{1\dots1}\right)
\ee
which reduces to a bell pair in the case $i=2$ and a GHZ state in the case $i=3$. For two parties the entanglement entropy is the unique measure of entanglement which define various LOCC classes (the states which are equivalent to one another under LOCC).

As soon as one considers more than two parties the situation becomes substantially more complex. Even asymptotically under the less restrictive stochastic-LOCC\footnote{SLOCC is less restrictive because it allows for the conversion protocol to be result dependent. That is, different operators may be applied depending on the outcome of the previous step.} (SLOCC) beyond three qubits there are a large number of nonequivalent forms of multipartite entanglement which are not well classified and grow quickly as the number of parties increase \cite{dur,bennett,walter}. As such in general it is not clear how to pick a target state for a multipartite distillation that would fully capture the entanglement structure of the state.

In order to connect our work here we make the following observation: given a cat state the entanglement entropy of any partition of the involved parties is 1. That is
\be
S(A_i)=S(A_iA_j)=\cdots=S(A_i...A_{n-1})=1.
\ee
In fact the cat-$k$ state can be represented as a simple hypergraph with a single $k$-edge with weight one. On this graph a single $k$-thread can be placed which constitutes the unique maximal hyperthread configuration. In analogy with the case of 2-threads it is very tempting to assert that a $k$-thread should be thought of as a distilled cat-$k$ state between the regions it connects. If this is true it leads to a number of interesting conclusions based on our results of maximal hyperthread configurations. Given a holographic CFT state:
\begin{outline}
\1 There exists a quantum channel which preserves the entanglement entropy of each region and distills the state to a fixed number of cat-2 and cat-$k$ states. The number of each is determined by the areas of $m_2$ and $t_k$
\be
\ket{\psi}\longrightarrow \ket{2}^{\otimes\left( \frac{1}{2}\area(m_2)-\frac{k}{\alpha}\area(t_k)\right)}\otimes\ket{k}^{\otimes \frac{1}{\alpha}\area(t_k)}
\ee
\1 There exists a quantum channel which distills the state into a fixed number of cat-$k$ states which is given by $\frac{1}{k}HP_{k}(\mathcal{A})=\frac{1}{\alpha}\area(m_k)$ which is strictly less than $\area(m_2)$. The loss of entanglement can be viewed as the consumption of bell pairs using quantum teleportation to manufacture the necessary correlations \cite{bone}. In general a large number of bell pairs are needed for each additional cat-$k$ produced beyond $\area(t_k)$
\be
\ket{\psi}\longrightarrow  \ket{k}^{\otimes \frac{1}{\alpha}\area(m_k)}.
\ee
\end{outline}
In other words this would mean the hyperthread partition $HP_{k}(\mathcal{A})$ is a measure of the distillable cat-$k$ entanglement of the CFT state.

It is well known that holographic spacetimes satisfy the monogamy of mutual information (MMI) \cite{2013PhRvD..87d6003H}
 \be 
 I_3(A:B:C) \coloneqq S(A)+S(B)+S(C)+S(ABC)-S(AB)-S(AC)-S(BC) \leq 0.
 \ee
This has been used to argue that holographic states can not contain large amounts of cat state entanglement as such states violate the inequality. Our position here is \emph{not} that this is false, but rather that if hyperthreads are distilled cat states then it may be possible that holographic states can be distilled to states which do. Note that such a distillation \emph{only} preserves the single party entropies. Since a hyperthread may cross surfaces more than once generically other RT surfaces (such as e.g. $m_{AB}$) will not be maximally saturated. As such the two party entropies of the distillation will generically be smaller leading to a potential violation of MMI. In fact direct calculation shows for four boundary regions
\be
-I_3\geq \mu^*(H_4)
\ee
where one should think of each possible hyperthread configuration providing a separate bound on the entanglement of the CFT state. Since a maximal configuration can always be constructed with \emph{only} 2-threads the tightest possible bound is given precisely by MMI.

The various dual programs defined in this paper are unlikely to provide a complete characterize the multipartite entanglement of a holographic state. Still, they seem to provide a compelling connection between geometry and information beyond the RT formula.  We view this as forward progress towards the challenging problem of a more complete and robust characterization of the multipartite entanglement of holographic CFT states in the context of quantum information theory.
 
We briefly mention that the methods used in this paper can be easily generalized to create possible hyperthread equivalents of other simple multipartite states (e.g. W states, perfect tensor states) by matching the contributions of a hyperthread to various entanglement entropies to that of the state. This may allow one to consider measurements and distillation protocols of a wider class of target states. We leave this to future work.

\subsection{MMI violating ``geometries"}

Consider a constant time slice of AdS$_3$ and split the boundary into four equal regions. Calculating the minimal surface homologous to each region, we define the equivalence relation
 \be
 \sim \coloneqq m_A \sim m_B \sim m_C \sim m_D
 \ee
\begin{figure}[H]
\begin{tabular}{cc}
\centering
\includegraphics[width=.45\textwidth,page=9]{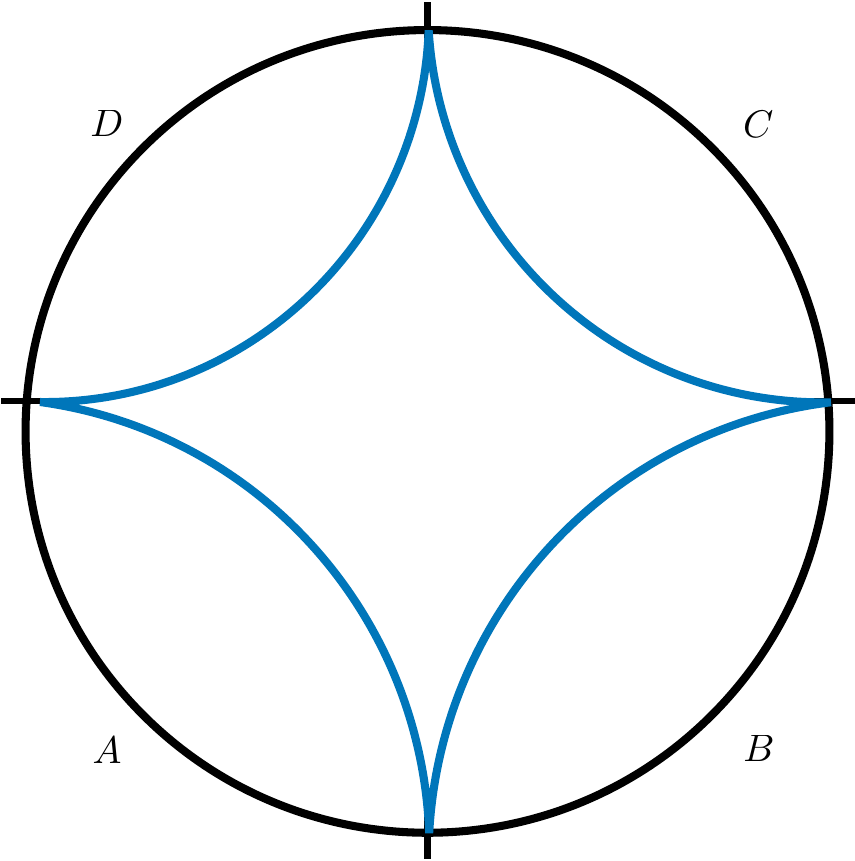}&
\includegraphics[width=.45\textwidth,page=11]{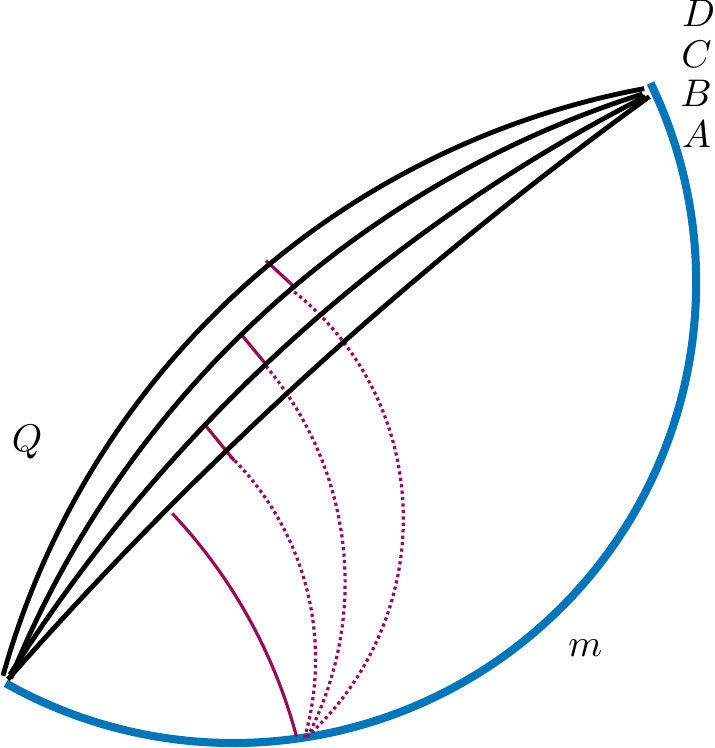}
\end{tabular}
\caption{\label{fig:Qspace} The space $Q$ is formed by the identification of the surfaces $m(A_i)$. A hyperthread on $Q$ can split on the minimal surface $m$ such that it connects to each boundary component. This allows for large amounts of higher party entanglement. The surface $m$ can be thought of as the generalization of a hyperedge.}
\end{figure}

 and from it create the quotient space $Q$ (see Figure \ref{fig:Qspace})
 \be 
 Q \coloneqq \mathbb{H}^2 \setminus \sim.
 \ee
 Note that the resulting space is \emph{not} a manifold nor an orbifold. However, each copy of the original homology region is a Riemannian manifold of codimension-0. While the surface $m$ is a codimension-1 Riemannian manifold. $Q$ falls into the classification of a stratifold \cite{wiki:111}: a generalization of a manifold which allows for such singularities.
 
 We will assume that there is a CFT state dual to such an object and that the RT formula holds analogously. Applying the RT formula to $Q$ we can calculate the entanglement entropies between a single (pair of) boundary region(s) and their complement. In all cases we have 
 \be
 S(A_i)=S(A_iA_j)=\area(m).
 \ee 
 This is important as it directly leads to a violation of MMI: since all the entropies are the same we have
 \be
 I_3 = \area(m) \geq 0
 \ee
 meaning the dual CFT state would contain significant amounts of higher party entanglement. This is easily seen using 4-threads which split on $m$ so that every hyperthread attaches to each boundary region. Importantly, this can be accomplished without \emph{any} 2-threads.
 
 Removing $m$ by a cut simultaneously separates $A,B,C,D$ and thus acts analogously to a hyperedge of a hypergraph. The values of the entanglement entropy can be reproduced as the cuts of a hypergraph with as single 4-edge in much the same way the entanglement entropy of CFT states can be modeled by bivalent graphs.
 
 It would be interesting to see if the assumptions made here hold and such stratifolds can be proven to be dual to a class of non-holographic CFT states which violate MMI.
 
\subsection{Signed threads}
In the optimization programs discussed in this paper the positivity of the metric over hyperthreads is a consequence of the requirement that in the dual each hyperthreads sees \emph{at least} the minimal required barrier. This is because of the duality which requires the corresponding Lagrange multiplier (dual variable) to take strictly positive values. If we change this constraint to an equality constraint, that is each thread sees \emph{exactly} the minimal required barrier, this is equivalent in the dual to removing the positivity constraint and allowing the measure to take both positive and negative values \cite{tao2011an,wiki:112}. That is we have for example the program
\be
\begin{split}
&\max\; k\mu(H_k)\text{ s.t. } \forall x\in \Sigma,\; \int_{H}d\mu(h)\Delta(x,h)\leq 1\\
&\min \nu(x)\text{ s.t. } \forall h\in H_k,\; \int_{\Sigma}d\nu(x)\Delta(x,h) =k
\end{split}
\ee
where $\mu$ can take either sign.

The Hahn decomposition theorem states that given such a signed measure it is always possible to decompose the set into a positive and negative part
\be
H = H^+ \cup H^-
\ee
on which we have separately the (non-signed) measures $\mu^+$ and $\mu^-$. The measure on $H$ is then given by
\be
\mu=\mu^+-\mu^-.
\ee
\begin{figure}[H]
\centering
\includegraphics[width=.5\textwidth,page=16]{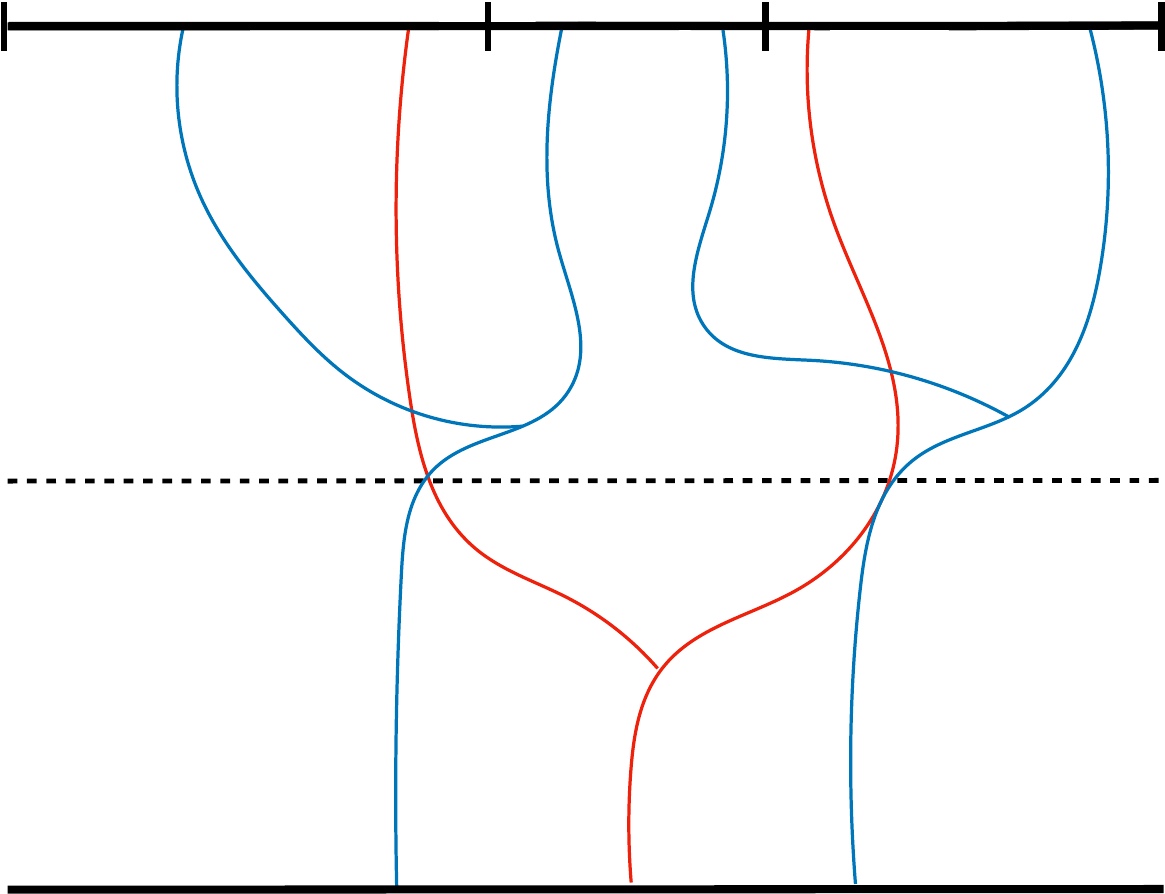}
\caption{\label{fig:signedthread} A negative 3-thread (red) allows additional positive 3-threads (blue) across a potential bottleneck (dashed line). These are arranged such that the total density of the threads on this surface is zero. Together they would contribute $6-3=3$ to an objective. This is made possible because the hyperthreads can split in different regions of the bulk. }
\end{figure}
\noindent The consequence of this decomposition is that each hyperthread in $H^+$ has a corresponding hyperthread in $H^-$ which contributes negatively to the objective and density bound. This creates a new mechanism where groups of positive and negative hyperthreads can be used to contribute to an objective without contributing to the saturation of threads on a potential bottleneck (see Figure \ref{fig:signedthread}). Note that the contributions of a positive and negative 2-thread cancel one another so this is a novel feature of hyperthreads $k\geq3$.

Considering such a minimization program, the only way \emph{all} hyperthreads can cross the same minimal required barrier is if they all split in the same bulk region. But, since this must hold for any location the hyperthreads can split, the only valid surface one could pick is the boundary regions being considered. As such, the mechanism described is in fact powerful enough to allow, in the case of vacuum AdS, for an infinite number of hyperthreads to be placed on the manifold due to the uv cutoff.

It is possible that negative hyperthreads may play a role in the calculation of entanglement measures such as those which can be written as a linear superposition of terms, some of which are negative.\footnote{This is very generic. For example all of the entropy inequalities which are facets of the holographic entropy cone \cite{EC} can be written in this manner. Subtracting all of the terms to one side gives a strictly positive quantity which one could ask for a dual thread description. Both the mutual information coming from subadditivity and tripartite information coming from MMI are examples.} Though to limit their effect the corresponding optimization programs will necessarily include additional constraints.

\acknowledgments
The work of J.H. is supported by the Simons Foundation through \emph{It from Qubit: Simons Collaboration on Quantum Fields, Gravity, and Information}. J.H would like to thank Chris Akers, Matt Headrick, and Brian Swingle for interesting and stimulating discussions and Ning Bao for comments on an early draft. J.H would like to especially thank Matt Headrick and Veronika Hubeny for sharing an early draft of a paper including the description of bit thread configurations as measurable sets. J.H. would like to thank Veronika Hubeny, Tom Faulkner, and the other organizers of TASI 2021 for hospitality during the later stages of this work.

\bibliographystyle{JHEP}
\bibliography{thread_distributions}
\end{document}